# Wavelet algorithm for the identification of P300 ERP component

© 2016, S.N. Agapov [1], V.A. Bulanov [2], A.V. Zaharov [3], M.S. Sergeeva [4].

[1] sergeyagapov@it-universe.ru
[2] vb@it-universe.ru
IT Universe LLC, Department of BCI. Samara, Russia.

[3] zakharov1977@mail.ru
Samara State Medical University, Department of Neurology and Neurosurgery. Samara, Russia.

[4] marsergr@yandex.ru
Samara State Medical University, Department of the Normal Physiology. Samara, Russia.

*Abstract:* Brain computer interfaces have many algorithms based on the P300 component of ERP. Modern industry has started to produce consumer grade EEG equipment which is handy and not too expensive. This gives us an opportunity to use BCI in everyday practice. In order to improve the performance of these devices we need effective algorithms for time series analysis and pattern recognition. We have tested Emotiv Insight headset in real live environment and we have conducted several tests for series of standard wavelets in P300 pattern recognition task.

*Competing interests*: The authors have declared that no competing interests exist.

# 1. Introduction

Modern population is becoming older every year. In the 20th century the average life expectancy rose significantly [1]. This resulted in bringing about several ageing-related blocking diseases such as amiotrophic lateral sclerosis (ALS), brainstem stroke and others [2, 3]. Heart diseases have become the main cause of death since the middle of the 20th century [4]. They lead to after-effects which can be the cause of complete or partial paralysis. The rehabilitation of these people increases the expenses allocated by a developed country's government on healthcare system [5, 6]. All these factors require fast development of effective and cheap brain-computer interfaces. If BCI equipment is used together with appropriate medical and rehab devices, it can considerably improve the abilities of movement-constrained patients.

Nowadays brain-computer interfaces (BCI) can use several different types of input signals from brain: the P300 component of ERP-wave, imagery movements, slow cortical potentials (SCP), steady state visually evoked potential (SSVEP) [7]. Modern industry has produced a lot of devices which use some of these concepts. However, in everyday rehabilitation clinical practice their use is limited. One of the main causes is the cost of the software and hardware parts of the brain-computer interfaces. Low-cost devices often have poor SNR (signal-to-noise ratio) characteristics. The modern methods of digital signal processing and machine learning can greatly improve the accuracy of consumer class equipment. This improvement should give an opportunity to use EEG devices for medical and rehabilitation facilities.

The implementation of wavelet analysis for the recognition of P300 ERP component in EEG signal registered with the help of Emotiv Insight (Emotiv Inc., http://emotiv.com/insight/) is shown in this paper.

# 2. Materials and methods

A group of 5 healthy men aged 29-44 (35.8 ± 7.2) participated in our experiments. We recorded EEG signals with 5-channel Emotive Insight EEG headset (Fig. 1). This device uses channels AF3, AF4, T7, T8, Pz according to the international 10-20 system (Fig. 2). The reference electrode was always placed at the left ear. We used software that showed stimulus on the computer display («eSpeller», Java 1.8, Windows 7, granted by CPR LLC, Samara, Russia). That software had been designed with "single character paradigm" [11].

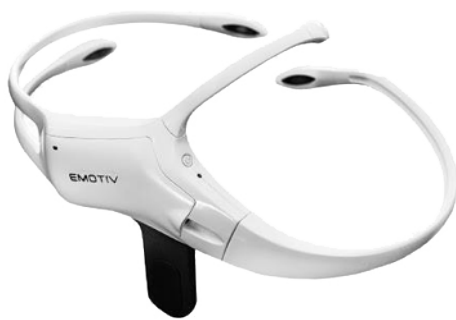

Fig. 1. Emotiv Insight 5-channel wireless EEG headset.

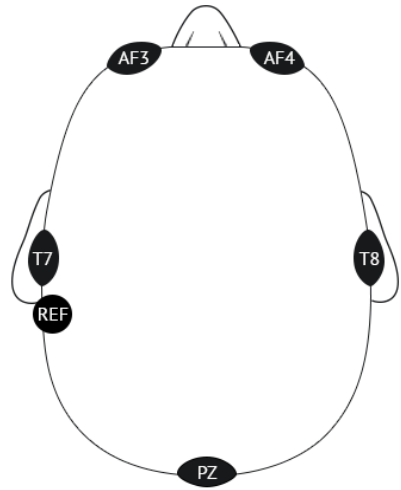

Fig. 2. Electrodes disposition.

EEG was recorded in a small office room; all tests were performed when it was daylight. There was some noise made by PC coolers. Datasets were recorded with the sampling rate 128 sec$^{-1}$. The obtained data was analyzed with MathWork® MATLAB R2015a (http://www.mathworks.com) in Department of BCI of Samara State Medical University, and R statistical package (https://www.r-project.org).

**2.1. Experimental design**

We carried out our experiments within 9 days between 12:00 and 18:00. Each participant sat in front of the PC display. The distance from a participant's eyes to the display was 0.5-0.7 m.

The participant wearing Emotiv Insight EEG headset was gazing at the display, when eSpeller software showed him the square window with 250-mm-side size divided by 9 equal cells in the form of 3x3 matrix with a dot in each cell. The software flashed the cells in random order. 'Flashed' means changed the image in the cell: it made the dot one and a half as large and it also altered the color from grey [RGB 179,179,179] to red [RGB 255, 0, 0] (Fig. 3). Each of the 9 cells was flashed only once in each cycle. One cycle consists of 9 flashes. We set stimulus duration to 120 ms and pause duration to 180 ms. Therefore, ISI (interstimulus interval) was set to 300 ms [22]. The 20 flashing cycles completed one experimental session. There was a 10-sec break between the sessions.

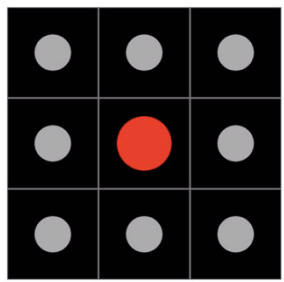

Fig. 3. The window of eSpeller.
Cells 1 – 4, 6 – 9 are not flashing, the red dot is flashing in cell 5.

Each participant had a task to concentrate on the only one cell at the moment. That cell was defined as *target cell* and the red dot in this cell was called *target visual stimulus*. Therefore, the other 8 cells were called *non-target cells*, and the red dots in them were named *non-target visual stimulus*. The cells were flashed in random order.

In the course of the experiment each participant gazed at cells 1-9 in succession (Fig. 4).

| 1 | 2 | 3 |
|---|---|---|
| 4 | 5 | 6 |
| 7 | 8 | 9 |

Fig. 4. The succession in which cells are gazed at.
A participant gazes at the cell 1 first, after 20 flashing cycles they have a 10-sec break and then they continue with cell 2. They repeat this procedure up to cell 9.

We presented visual stimuli and recorded EEG signals making time stamps simultaneously. As a result, we got 9 files, one for each session, in accordance with the number of cells which were gazed at by a participant. We conducted 3 trials for each participant. All the participants were healthy dexter men and they were volunteers. In this paper the *subjA, subjB, subjK, subjP, subjS* represent the participants. Therefore, we conducted 15 trials and recorded 135 EEG-signals – 9 files × 3 trials × 5 participants.

**2.2. Continuous and discrete wavelet transformations**

The Fourier transform requires stationarity of the signals and does not locate any events in the signal in time domain. The wavelets give us this opportunity. The wavelets spread in many areas such as time series analysis, image processing, variance stabilization and others. [12, 13]. One of the features of the wavelets is scaling, which variates the width of the time window used for the analysis [14]. A wide class of functions can be represented as wavelets, and then we can create an adapted wavelet from time series data. Up until now many different wavelets have been created and they have a wide area of applications. For example, they are Morlet, Daubechies, "mexican hat" wavelets. The Matlab wavelet toolbox presents 16 distinct wavelet families.

The function $\psi(t)$ which satisfies the next conditions is called wavelet:

$$\int_{-\infty}^{\infty} \psi(u)du = 0, \quad \int_{-\infty}^{\infty} \psi^2(u)du = 1$$

When these conditions are adhered to, there should be an interval $[-T, T]$ of finite length and the value of $\psi(t)$ close to 0 outside of that interval. As a result, we obtain the "small wave" or "wavelet" [15]. We used Daubechies wavelets (db4 and db6) for our work, which are shown in Fig. 5.

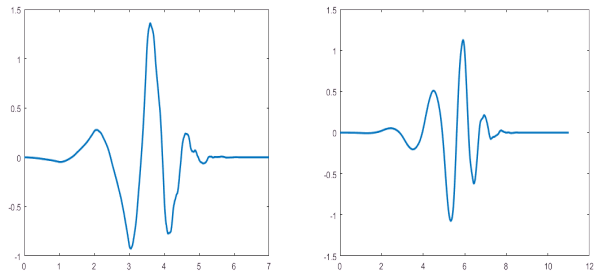

Fig 5. Daubechies wavelets: db4 (left), db6 (right).

The continuous wavelet transform (CWT) of a signal $x(t) \in L^2 R$ is defined as a dot product between the signal and the wavelet functions $\psi_{a,b}(t)$:

$$C_{a,b} = <x(t), \psi_{a,b}(t)>$$

$C_{a,b}$     the coefficients of the wavelet,
$\psi_{a,b}(t)$     scaled and translated wavelet function $\psi(t)$:

$$\psi_{a,b}(t) = \sqrt{|a|}\, \psi((t-b)/a)$$

where   $a$ — scale,
         $b$ — translation.

The CWT gives coefficients $C_{a,b}$ which have maximum values at these scales and times where the signal form is similar to that of the wavelet function $\psi(t)$. We make CWT with Matlab cwt() function.

The discrete wavelet transformation (DWT) is defined for discrete scale and translation parameters $\{a_j = 2^j,\ b_{j,k} = 2^j k\}$. These parameters give us a discrete family of wavelets:

$$\psi_{j,k}(t) = 2^{-j}\, \psi(2^{-j} t - k), \quad j, k \in Z$$

Therefore, we can define discrete (dyadic) wavelet transform as:

$$C_{j,k} = <x(t), \psi_{j,k}(t)>$$

The discrete wavelet transform adds some more constraints to the function $\psi(t)$ [16, 17]. However, we can still perfectly reconstruct a signal $x(t)$ from its discrete wavelet transform coefficients. In the case of the orthogonal wavelets this is:

$$x(t) = \sum_{j,k} C_{j,k}\, \psi_{j,k}(t), \quad j, k \in Z$$

# 3. Results

The CWT can be used for the recognition of the patterns which have the form of the wavelet function [18]. When we transform a signal to the wavelet coefficients, the maximum coefficient gives time and scale where the pattern appears. We have analyzed the signals which were recorded during experiment using well-known wavelets such as Daubechies (db 4, 6), reverse biorthogonal wavelets (rbio) and symlets (sym). These wavelets were used in many previous works for EEG filtering and analysis [19 - 21].

One epoch consisted of 128 samples (1 sec), which were counted from the start of visual stimulus. Then we averaged 20 epochs and next we eliminated all linear trends and normalized the averaged signal. For each wavelet we filtered a signal with the same wavelet. For the CWT we used scales 1-64, which covered frequencies 0-30 Hz, and the time interval from 230 ms to 700 ms. It is known from the previous works that P300 component normally appears in that time window [22]. We calculated the total of coefficient's maximum at each time point; this statistic was used for pattern recognition:

$$S = \sum_{b=1}^{w} \max(C_{a,b})$$

where $C_{a,b}$ CWT coefficients,
$w$ - time window width.

We obtained the accuracy of pattern recognition of P300 component for given wavelets. In Tables 1 - 4 we showed the accuracy of the algorithm for different numbers of the averaged epochs. The maximum accuracy for each participant is marked yellow. The maximum accuracy among the participants is marked green.

Table 1. P300 detection accuracy (%). The number of the averaged epochs is 5.

| Participant | Wavelet | | | | |
| --- | --- | --- | --- | --- | --- |
| | *Db4* | *Db5* | *Db6* | *Sym6* | *Rbio3.5* |
| *subjA* | 14.81 | 11.11 | 18.52 | 14.81 | 14.81 |
| *subjB* | 44.44 | 44.44 | 44.44 | 40.74 | 37.04 |
| *subjK* | 22.22 | 18.52 | 14.81 | 18.52 | 18.52 |
| *subjP* | 25.93 | 40.74 | 33.33 | 25.93 | 33.33 |
| *subjS* | 40.74 | 44.44 | 44.44 | 44.44 | 48.15 |

Table 2. P300 detection accuracy (%). The number of the averaged epochs is 10.

| Participant | Wavelet | | | | |
| --- | --- | --- | --- | --- | --- |
| | *Db4* | *Db5* | *Db6* | *Sym6* | *Rbio3.5* |
| *subjA* | 29.63 | 22.22 | 18.52 | 25.93 | 22.22 |
| *subjB* | 37.04 | 33.33 | 22.22 | 33.33 | 25.93 |
| *subjK* | 37.04 | 37.04 | 25.93 | 37.04 | 37.04 |
| *subjP* | 33.33 | 44.44 | 40.74 | 33.33 | 29.63 |
| *subjS* | 51.85 | 55.56 | 55.56 | 55.56 | 55.56 |

Table 3. P300 detection accuracy (%). The number of the averaged epochs is 15.

| Participant | Wavelet | | | | |
|---|---|---|---|---|---|
| | Db4 | Db5 | Db6 | Sym6 | Rbio3.5 |
| subjA | 29.63 | 29.63 | 33.33 | 48.15 | 22.22 |
| subjB | 44.44 | 51.85 | 40.74 | 44.44 | 40.74 |
| subjK | 44.44 | 40.74 | 37.04 | 44.44 | 37.04 |
| subjP | 22.22 | 37.04 | 37.04 | 33.33 | 29.63 |
| subjS | 59.26 | 70.37 | 66.67 | 59.26 | 70.37 |

Table 4. P300 detection accuracy (%). The number of the averaged epochs is 20.

| Particiapnt | Wavelet | | | | |
|---|---|---|---|---|---|
| | Db4 | Db5 | Db6 | Sym6 | Rbio3.5 |
| subjA | 55.56 | 44.44 | 48.15 | 51.85 | 40.74 |
| subjB | 62.96 | 66.67 | 62.96 | 59.26 | 62.96 |
| subjK | 51.85 | 51.85 | 44.44 | 51.85 | 33.33 |
| subjP | 33.33 | 37.04 | 37.04 | 37.04 | 37.04 |
| subjS | 77.78 | 77.78 | 77.78 | 74.07 | 81.48 |

# 4. Discussion

We can sum up the results obtained from our experiments:

- classification accuracy depends strongly on the number of target stimuli. For example, the maximum accuracy for participant *subjS* changed from 48,15% to 81,48%, when we increased the number of target stimuli from 5 to 20;

- classification accuracy strongly depends on a participant. For example, the accuracy for participant *subjP* was 37,04% and 81,48% for participant *subjS* (within 20 averaged epochs for both)**;**

- classification accuracy depends weakly on the wavelet chosen. For example, the accuracy for participant *subjS* was 74.07% with wavelet Sym6 and 81,48% with wavelet Rbio3.5;

- the average of the classification accuracy was 54,4±7,7% with 20 target stimuli. The minimum was 33,33% for participant *subjP* with wavelet Db4; the maximum was 81,48% for participant *subjS* with wavelet Rbio3.5.

We can conclude that the current realization of wavelet algorithm for patter recognition in EEG signals has several drawbacks: a low level of generalization and low overall accuracy even with a large quantity of target stimuli. We think there are several reasons for this:

- there are anthropomorphic differences between subjects. The shape and time localization of the ERP-waves and their P300 component have high variance among humans. Our results confirm that well-known fact;

- the wavelets chosen, in general, have a shape similar to ERP-wave, but this is not a sufficient condition for detection. We can suppose that the adapted wavelet will show a better performance;

- we can presume that the Emotiv Insight device has hardware or software design features which make an effect on the given algorithm and do not allow us to get a better result.

# 5. Conclusion

We have presented an algorithm for the P300 component detection in an EEG signal using several well-known wavelets. This algorithm can be used with Emotiv Insight device for a certain category of people. Additional testing of the accuracy level for a certain participant is required before using the given algorithm. Accuracy is likely to be improved if the adapted wavelet is used, and we are planning this investigation in the future.

The results obtained need more investigation. For example, we can test the given algorithm on the data acquired with other EEG equipment. Also, we could try to adapt a wavelet to the individual pattern of the ERP wave of a participant.

# References

[1] J. Oeppen, J. W. Vaupel, and others, "Broken limits to life expectancy," *Science (80-. ).*, vol. 296, no. 5570, pp. 1029–1031, 2002.

[2] G. Logroscino, R. Tortelli, G. Rizzo, B. Marin, P. M. Preux, and A. Malaspina, "Amyotrophic lateral sclerosis: An aging-related disease," *Curr. Geriatr. Reports*, vol. 4, no. 2, pp. 142–153, 2015.

[3] B. a Yankner, T. Lu, and P. Loerch, "The aging brain.," *Annu. Rev. Pathol.*, vol. 3, pp. 41–66, 2008.

[4] [Online]. Available: http://www.who.int/mediacentre/factsheets/fs310/ru/.

[5] K. McGregor and B. Pentland, "Head injury rehabilitation in the U.K.: An economic perspective," *Soc. Sci. Med.*, vol. 45, no. 2, pp. 295–303, 1997.

[6] C. D.N. and O. J., "A clinical and economic perspective on head injury rehabilitation," *Journal of Head Trauma Rehabilitation*, vol. 8, no. 4. pp. 1–14, 1993.

[7] G. Edlinger, B. Z. Allison, and C. Guger, "How many people can use a BCI system?," in *Clinical Systems Neuroscience*, 2015, pp. 33–66.

[8] R. Quian Quiroga, "Obtaining single stimulus evoked potentials with wavelet denoising," *Phys. D Nonlinear Phenom.*, vol. 145, no. 3–4, pp. 278–292, 2000.

[9] R. Quian Quiroga, O. W. Sakowitz, E. Basar, and M. Schürmann, "Wavelet Transform in the analysis of the frequency composition of evoked potentials," *Brain Res. Protoc.*, vol. 8, no. 1, pp. 16–24, 2001.

[10] X. Li, X. Chen, Y. Yan, W. Wei, and Z. Wang, "Classification of EEG Signals Using a Multiple Kernel Learning Support Vector Machine," *Sensors*, vol. 14, no. 7, pp. 12784–12802, 2014.

[11] R. Fazel-rezai, S. Gavett, W. Ahmad, A. Rabbi, and E. Schneider, "A Comparison among Several P300 Brain-Computer Interface Speller Paradigms," *Clin. EEG Neurosci.*, vol. 42, no. 4, pp. 209–213, 2011.

[12] C. Heil, "Ten Lectures on Wavelets (Ingrid Daubechies)," *SIAM Review*, vol. 35, no. 4. pp. 666–669, 1993.

[13] S. Mallat, "A Wavelet Tour of Signal Processing," *A Wavelet Tour Signal Process.*, pp. 20–41, 1999.

[14] C. K. Chui, *Wavelets : a tutorial in theory and applications*, no. 2. 1992.

[15] D. B. Percival and A. T. Walden, "Wavelet Methods for Time Series Analysis," *J. Nonparametr. Stat.*, vol. 10, no. December 2014, p. 594, 2000.

[16] E. Erçelebi, "Electrocardiogram signals de-noising using lifting-based discrete wavelet transform," *Comput. Biol. Med.*, vol. 34, no. 6, pp. 479–493, 2004.

[17] M. V. Wickerhauser, A. Jensen, A. la Cour-Harbo, A. Boggess, and F. J. Narcowich, *Ripples in Mathematics: The Discrete Wavelet Transform*, vol. 110, no. 2. 2003.

[18] D. B. Percival and A. T. Walden, *Wavelet Methods for Time Series Analysis*. 2000.

[19] A. B. Geva and D. H. Kerem, "Forecasting generalized epileptic seizures from the EEG signal by wavelet analysis and dynamic unsupervised fuzzy clustering," *IEEE Trans. Biomed. Eng.*, vol. 45, no. 10, pp. 1205–1216, 1998.

[20] V. J. Samar, A. Bopardikar, R. Rao, and K. Swartz, "Wavelet Analysis of Neuroelectric Waveforms: A Conceptual Tutorial," *Brain Lang.*, vol. 66, no. 1, pp. 7–60, 1999.

[21] M. Murugappan, "Classification of human emotion from EEG using discrete wavelet transform," *J. Biomed. Sci. Eng.*, vol. 03, no. 04, pp. 390–396, 2010.

[22] S. J. Luck, *An Introduction to the Event-Related Potential Technique*, vol. 78, no. 3. 2005.